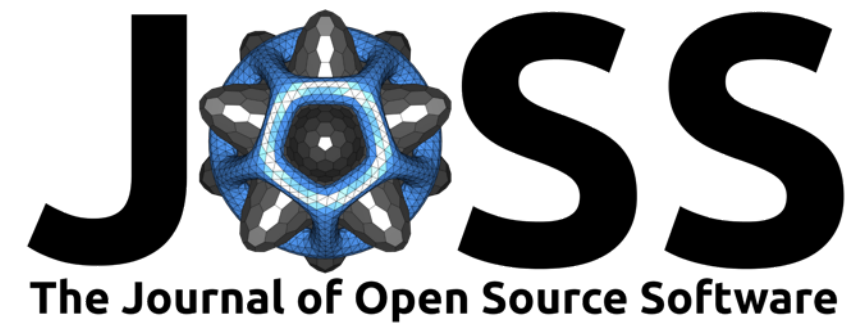

# JAXtronomy: A JAX port of lenstronomy

**Alan Huang** [1], **Simon Birrer** [1], **Natalie B. Hogg** [2], **Aymeric Galan** [3,4], **Daniel Gilman** [5], **Anowar J. Shajib** [5,6,7], and **Nan Zhang** [8]

**1** Department of Physics and Astronomy, Stony Brook University, Stony Brook, NY 1794, USA **2** Laboratoire Univers et Particules de Montpellier, CNRS and Université de Montpellier (UMR-5299), 34095 Montpellier, France **3** Max-Planck-Institut für Astrophysik, Karl-Schwarzschild Straße 1, 85748 Garching, Germany **4** Technical University of Munich, TUM School of Natural Sciences, Physics Department, James-Franck-Straße 1, 85748 Garching, Germany **5** Department of Astronomy and Astrophysics, University of Chicago, Chicago, IL 60637, USA **6** Kavli Institute for Cosmological Physics, University of Chicago, Chicago, IL 60637, USA **7** Center for Astronomy, Space Science and Astrophysics, Independent University, Bangladesh, Dhaka 1229, Bangladesh **8** Department of Physics, University of Illinois, 1110 West Green St., Urbana, IL 61801, USA

## Summary

Gravitational lensing is a phenomenon where light bends around massive objects, resulting in distorted images seen by an observer. Studying gravitationally lensed systems provides insights into cosmology and astrophysics, including constraints of the expansion rate of the Universe and the distribution of dark matter.

Thus, we introduce `JAXtronomy`, a re-implementation of the gravitational lensing software package `lenstronomy` (Birrer, 2021; Birrer & Amara, 2018) using JAX (Bradbury et al., 2018). JAX is a Python library that uses an accelerated linear algebra (XLA) compiler to improve the performance of computing software. Our core design principle of `JAXtronomy` is to maintain an identical API to that of `lenstronomy`.

The main JAX features utilized in `JAXtronomy` are just-in-time compilation, which can lead to significant reductions in execution time, and automatic differentiation, which allows for the implementation of gradient-based algorithms that were previously impossible. Additionally, JAX allows code to be run on GPUs or parallelized across CPU cores, further boosting the performance of `JAXtronomy`.

## Statement of need

`lenstronomy` has been widely applied to numerous science cases, with more than 200 publications making use of the software, and has an increasing number of dependent packages relying on features of `lenstronomy`. For instance, science cases directly involving `lenstronomy` include galaxy evolution studies using strong lensing (Shajib et al., 2021; Sheu et al., 2025; Tan et al., 2024) and detailed lens modeling for measuring the Hubble constant using time-delay cosmography by the TDCOSMO collaboration (Birrer et al., 2020, 2025; Birrer & Treu, 2021; Gilman et al., 2020; Millon et al., 2020; Schmidt et al., 2025; Shajib et al., 2022; Williams et al., 2025).

Examples of packages dependent on `lenstronomy` for general-purpose lensing computations and image modelling include the `dolphin` package (Shajib et al., 2025) for automated lens modeling, the `galight` package (Ding et al., 2020) for galaxy morphology measurements, SLSim (Khadka et al, 2026, in prep) for simulating large populations of strong lenses, pyHalo (Gilman et al., 2019) and `mejiro` (Wedig et al., 2025) for simulating strong lenses with dark

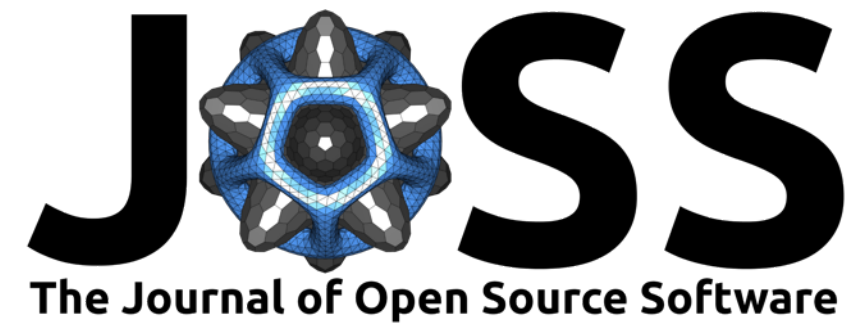

matter substructure, and paltas (Wagner-Carena et al., 2023) for neural network inference tasks.

In many of these applications, computational constraints are the key limiting factor for strong gravitational lensing science. For example, increased data quality and number of lenses to analyze makes lens modeling a computational bottleneck, and expensive ray-tracing through tens of thousands of dark matter substructures limits the amount of images that can be simulated, especially for the training of neural networks and simulation-based inferences. These ever-increasing computational costs have led to the development of several JAX-accelerated and GPU-accelerated strong-lensing packages, such as GIGALens (Gu et al., 2022), Herculens (Vernardos et al., 2022), paltax (Wagner-Carena et al., 2024), GLaD (Wang et al., 2025), Caustics (Stone et al., 2024) and Google Research's jaxstronomy (Google Research, 2023).

### Why JAXtronomy?

JAXtronomy inherits a wide range of features from lenstronomy that are not offered by any of the aforementioned JAX-accelerated or GPU-accelerated software. These features include lenstronomy's linear amplitude solver, which reduces the number of sampled parameters during lens modeling, as well as a variety of log-likelihood functions and optional punishment terms to improve robustness during fitting. JAXtronomy aims to maintain an identical API to lenstronomy so that packages dependent on lenstronomy can transition seamlessly to JAXtronomy.

## Improvements over lenstronomy in image simulation

The simulation of a lensed image comes in three main steps. The first step begins with a coordinate grid in the angles seen by the observer. These coordinates are ray-traced through the deflectors back to the source plane. This process requires the calculation of light-ray deflection angles at each deflector. Second, the surface brightness of the source is calculated on the ray-traced coordinate grid. This produces a lensed image. Third, the lensed image gets convolved by the point spread function (PSF) originating from diffraction of the telescope optics and atmospheric turbulence. Due to the various choices in deflector mass profiles, light model profiles, grid size, and PSF kernel size, the overall runtime of the pipeline can vary significantly.

In the following sections, we outline the improvements in performance that JAXtronomy has over lenstronomy for each step in the pipeline. These performance benchmarks were run using an Intel(R) Xeon(R) Gold 6338 CPU @ 2.00GHz, an NVIDIA A100 GPU, and JAX version 0.7.0.

### Deflection angle calculations

Each entry in the table indicates how much faster JAXtronomy is compared to lenstronomy at computing deflection angles for the corresponding deflector profile and grid size. Some comparisons vary significantly with values of function arguments, so a range is given rather than a number.

| Deflector profile | 60x60 grid (CPU) | 180x180 grid (CPU) | 180x180 grid (GPU) |
|---|---|---|---|
| CONVERGENCE | 0.4x | 1.1x | 0.5x |
| CSE | 1.6x | 2.6x | 2.6x |
| EPL | 5.1x - 15x | 9.2x - 17x | 37x - 120x |
| EPL (JAX) vs EPL_NUMBA | 1.4x | 3.0x | 13x |
| EPL_MULTIPOLE_M1M3M4 | 2.1x - 7x | 6.4x - 13x | 42x - 108x |
| HERNQUIST | 2.0x | 3.4x | 5.8x |

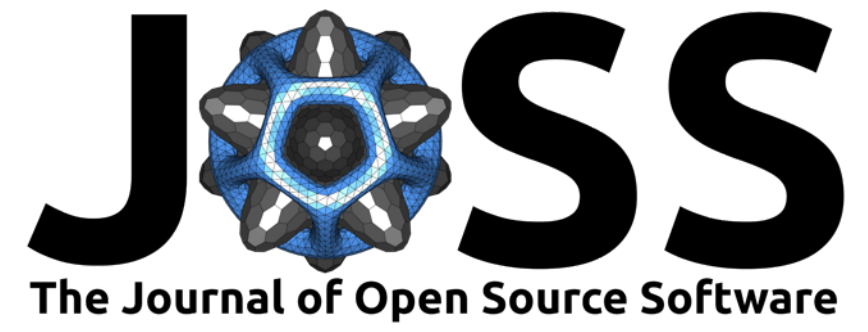

| Deflector profile | 60x60 grid (CPU) | 180x180 grid (CPU) | 180x180 grid (GPU) |
| --- | --- | --- | --- |
| HERNQUIST_ELLIPSE_CSE | 3.8x | 5.4x | 40x |
| MULTIPOLE | 0.9x | 1.0x | 8.3x - 14x |
| MULTIPOLE_ELL | 1.5x - 2.1x | 2.0x - 2.8x | 70x |
| NIE/SIE | 0.5x | 0.5x | 2.0x |
| NFW | 1.6x | 3.3x | 4.5x |
| NFW_ELLIPSE_CSE | 4.1x | 6.7x | 31x |
| PJAFFE | 1.0x | 1.2x | 2.8x |
| PJAFFE_ELLIPSE_POTENTIAL | 1.4x | 1.6x | 3.1x |
| SHEAR | 0.7x | 2.0x | 0.9x |
| SIS | 1.4x | 3.3x | 2.0x |
| TNFW | 2.4x | 5.8x | 7.5x |

Due to JAX's higher function call overheads compared to standard Python, slowdowns can occur when using computationally simple deflector profiles. These deflector profiles benefit most from the parallelization that JAX offers.

## Flux calculations

An analogous table for the different light profiles is shown below. The MULTI_GAUSSIAN and MULTI_GAUSSIAN_ELLIPSE profiles include five GAUSSIAN and GAUSSIAN_ELLIPSE components, respectively, highlighting JAX's improved performance in sequential computations.

| Light profile | 60x60 grid (CPU) | 180x180 grid (CPU) | 180x180 grid (GPU) |
| --- | --- | --- | --- |
| CORE_SERSIC | 2.0x | 6.7x | 4.2x |
| GAUSSIAN | 1.0x | 2.5x | 1.3x |
| GAUSSIAN_ELLIPSE | 1.5x | 3.6x | 2.0x |
| MULTI_GAUSSIAN | 3.7x | 11x | 7.8x |
| MULTI_GAUSSIAN_ELLIPSE | 4.0x | 13x | 6.9x |
| SERSIC | 1.0x | 1.7x | 3.9x |
| SERSIC_ELLIPSE | 1.9x | 5.7x | 3.2x |
| SERSIC_ELLIPSE_Q_PHI | 1.7x | 5.5x | 3.3x |
| SHAPELETS (n_max=6) | 6.2x | 3.4x | 15x |
| SHAPELETS (n_max=10) | 6.0x | 4.5x | 17x |

## FFT convolution

We find that FFT convolution using JAX on CPU results in variable performance boosts or slowdowns compared to `lenstronomy` (which uses SciPy's FFT convolution). On a 60x60 grid, and kernel sizes ranging from 3 to 45, JAX on CPU ranges from being 1.1x to 2.9x faster than `lenstronomy`, with no obvious correlation to kernel size. On a 180x180 grid, and kernel sizes ranging from 9 to 135, `JAXtronomy` on CPU ranges from being 0.7x to 2.5x as fast as `lenstronomy`, with no obvious correlation to kernel size.

However, FFT convolution using JAX on GPU is significantly faster than SciPy. On a 60x60 grid, and kernel sizes ranging from 3 to 45, JAX on GPU ranges from being 1.5x to 3.5x faster than `lenstronomy`, with JAX performing better at higher kernel sizes. On a 180x180 grid, and kernel sizes ranging from 9 to 135, `JAXtronomy` on GPU is about 10x to 20x as fast as `lenstronomy`, again with JAX performing better at higher kernel sizes.

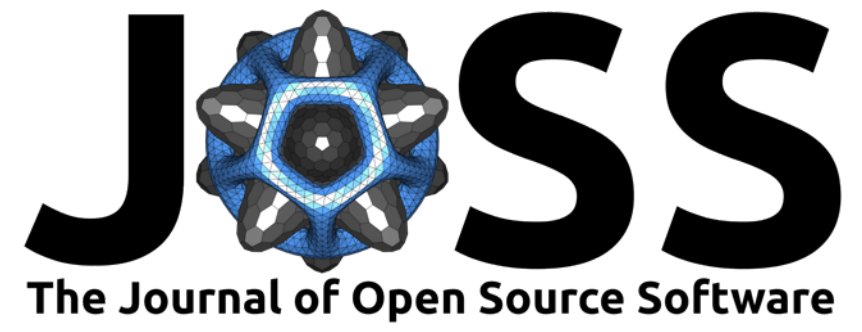

## Improvements over lenstronomy in lens modelling

The process of lens modelling involves finding best-fit parameters describing a lensed system from real data. In `lenstronomy`, this typically involves a Particle Swarm Optimizer (PSO, Kennedy & Eberhart, 1995) for optimization and Monte Carlo Markov Chains for posterior sampling. `JAXtronomy` retains these lens modelling algorithms from `lenstronomy` while benefitting from the increased performance outlined above.

In the following table, we compare `JAXtronomy`'s PSO performance to that of `lenstronomy` when modeling a lens with a singular isothermal ellipsoid (SIE) mass profile, Sersic-ellipse source and lens light profile, and a quadruply-imaged point source. The image is simulated using a 100x100 grid and FFT convolved using a PSF kernel with a size of 13 pixels.

| Device | 64 Particles | 128 Particles | 256 Particles | 512 Particles |
|---|---|---|---|---|
| lenstronomy (baseline) | 59s | 138s | 245s | 555s |
| 1 CPU core | 3x | 3x | 3x | 4x |
| 2 CPU cores | 5x | 6x | 6x | 7x |
| 4 CPU cores | 8x | 12x | 11x | 12x |
| 8 CPU cores | 11x | 17x | 17x | 24x |
| 16 CPU cores | 13x | 20x | 22x | 29x |
| 32 CPU cores | 13x | 20x | 20x | 29x |
| GPU | 8x | 7x | 27x | 46x |

Additionally, using JAX's autodifferentiation, we have implemented the L-BFGS gradient descent algorithm from the Optax library (Babuschkin et al., 2020) for optimization. This is a significant improvement over `lenstronomy`'s PSO, which does not have access to gradient information. Due to the stochastic nature of the PSO, we do not present a concrete comparison between `lenstronomy`'s PSO and `JAXtronomy`'s minimizer for the time it takes to find best-fit parameters.

## Acknowledgements

AH and SB are supported by DoE Grant DE-SC0026113, NASA Grants JWST-GO-07184 and 22-ROMAN22-0072. Major software dependencies of `JAXtronomy` not previously mentioned include NumPy (Harris et al., 2020), SciPy (Virtanen et al., 2020), and NumPyro (Bingham et al., 2019; Phan et al., 2019).

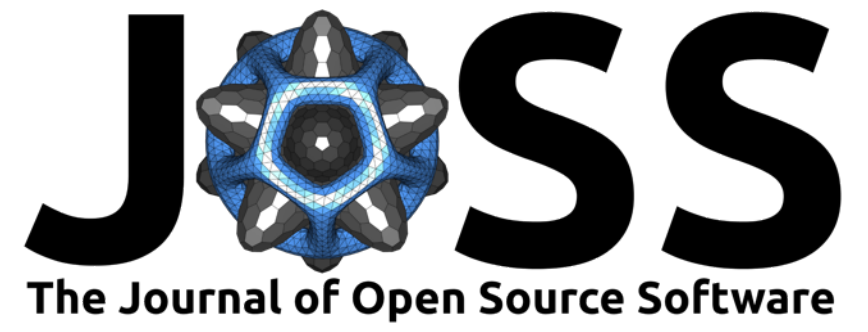

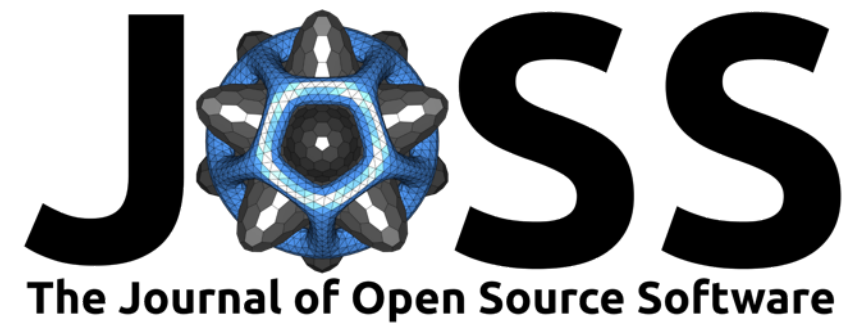

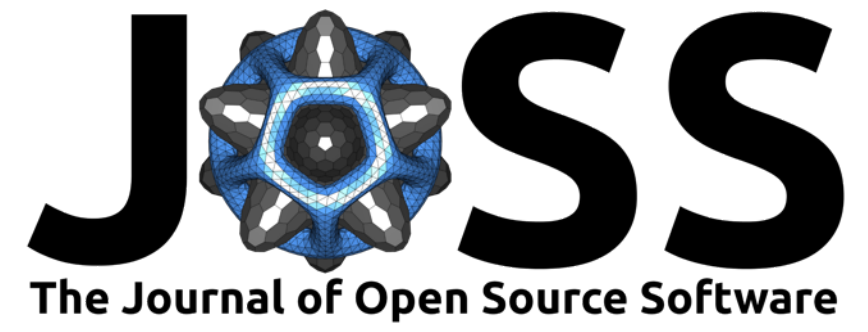